\documentclass[12pt, onecolumn ]{IEEEtran}

\usepackage{url}
\usepackage{amsmath}
\usepackage{amsfonts}
\usepackage{theorem}
\usepackage[dvips]{epsfig}      % ******* for figures ***************
\usepackage{graphicx}
\usepackage{enumerate}
\usepackage{color}                  %******** for colored remarks********

\usepackage{verbatim}

\usepackage{amssymb}
\usepackage{amsbsy}
\usepackage{setspace}
\usepackage{url}

\usepackage{float}

\theoremstyle{plain}

\newtheorem{Proposition}{Proposition}

\newtheorem{Problem}{Problem}
\newtheorem{Fact}{Fact}

 \newcommand{\Int}{\operatorname{int}}
 \newcommand{\real}{\operatorname{Re}}
 
 \newcommand{\diag}{\operatorname{diag}}

{\theorembodyfont{\rmfamily} }
{\theorembodyfont{\rmfamily} }
{\theorembodyfont{\rmfamily} \newtheorem{Example}{Example}}
{\theorembodyfont{\rmfamily} }
{\theorembodyfont{\rmfamily} }

\newcommand {\R}{\mathbb R}

\newcommand{\be}{\begin{equation}}
\newcommand{\ee}{\end{equation}}
\newcommand{\sgn}{\operatorname{{\mathrm sgn}}}
\newcommand{\LMD}{\lambda_0,\dots,\lambda_n}
\newcommand{\LMDRFMR}{\lambda_1,\dots,\lambda_n}

\newcommand{\LMDSTAR}{\lambda_0^*,\dots,\lambda_n^*}

\newcommand{\sname}{} \newcommand{\slabel}[1]{\debug{\fbox{\tiny \sname #1}}\label{\sname #1}}
\newcommand{\debug}[1]{}              % final version
\newcommand{\FB}{\begin{figure}[t]\centering} \newcommand{\FE}[2]{\caption{#2 \debug{\fbox{\sname #1}}} \slabel{#1} \end{figure}} \newcommand{\tB}{\begin{table}[hbtp]\centering}
\newcommand{\tE}[2]{\caption{#2 \debug{\fbox{\sname #1}}}\slabel{#1} \end{table}} 
\begin{document}
 \title{On the Ribosomal Density that Maximizes Protein Translation Rate\thanks{The research of MM and TT is partially supported by  research grants from  the Israeli Ministry of Science, Technology, and Space, and the
 Binational Science Foundation.
The research of MM is also supported by a research grant from the Israel Science Foundation}}
\author{ Yoram Zarai,   Michael Margaliot, and Tamir Tuller* \IEEEcompsocitemizethanks{
\IEEEcompsocthanksitem
 Y. Zarai is with the School of Elec. Eng., Tel-Aviv
University, Tel-Aviv 69978, Israel.
E-mail: yoramzar@mail.tau.ac.il
\IEEEcompsocthanksitem
M. Margaliot is with the School of Elec. Eng. and the Sagol School of Neuroscience, Tel-Aviv
University, Tel-Aviv 69978, Israel.
E-mail: michaelm@eng.tau.ac.il
\IEEEcompsocthanksitem
  T. Tuller (corresponding author) is with the Dept. of Biomedical Eng. and the Sagol School of Neuroscience, Tel-Aviv
University, Tel-Aviv 69978, Israel.
E-mail: tamirtul@post.tau.ac.il
 }}

\maketitle
%%%%%%%%%%%%%%%%%%%%%%%%%%%%%%%%%%%%%%%%%%%%%%%%%%%%%%%%%%

\begin{abstract}
%%%%%%%%%%%%%%%%%
During mRNA translation, several ribosomes   attach to the same mRNA molecule
simultaneously translating it   into a protein.
This  pipelining increases the protein production rate.
A natural and important question is what   ribosomal density   maximizes
the protein production rate. Using mathematical models
of ribosome flow along both a linear and a circular  mRNA molecule  we prove
   that typically the steady-state
production rate is maximized when the ribosomal density is one half of the maximal possible density.
We discuss the implications of our results to endogenous genes under
 natural cellular conditions and also to synthetic biology.
%%%%
%%%%%%
\end{abstract}
%%%%%%%%%%%%%%%%%%%%%%%%%%%%%%%%%%%%%%%%%%%%%%%%%%%%%%%%%%%%%%%%%%%%%%%%%%%%


%%%%%%%%%%%%%%%%%%%%%%%%%%%%%%%%%%%
\section{Introduction}
%%%%%%%%%%%%%%%%%%%%%%%%%%%%%%%%%%%
The transformation of the genetic information in the~DNA into functional proteins is called \emph{gene expression}. Two important steps in gene expression are   \emph{transcription} of the~DNA code into messenger~RNA (mRNA) by RNA polymerase~(RNAP), and then \emph{translation} of the~mRNA  into proteins. During   translation,  complex macromolecules called ribosomes traverse the mRNA strand, decoding it codon by codon into a corresponding chain of amino-acids that  is folded co- and post-translationally to become a functional protein~\cite{Alberts2002}. The rate in which proteins are produced during the translation step is called the protein translation rate or protein production rate.

According to   current knowledge, translation takes place
 in all living organisms and under all conditions. Understanding the numerous  factors that affect  this dynamical
process  has important implications to many scientific disciplines including medicine, evolutionary biology, synthetic biology, and more.

Computational models of translation   are becoming increasingly important as the amount of experimental findings  related to translation   rapidly increases (see, e.g. \cite{Zhang1994,Dana2011,Heinrich1980,MacDonald1968,TullerGB2011,Tuller2007,Chu2012,Shah2013,Deneke2013,Racle2013}). Such models are particularly important in the context of synthetic biology and biotechnology, as they can provide predictions on the qualitative and quantitative effects of  various manipulations of the genetic machinery. Recent advances in measuring translation in
\emph{real time}~\cite{Yan2016,Wu2016,Morisaki2016,ChongWang2016}
will  further increase the interest in computational models
 that can integrate and explain the measured biological data.

During translation, a large number of ribosomes act simultaneously on the same mRNA molecule.
This pipelining of the protein production leads to
 a more continuous production rate and increased  efficiency.
Indeed, the production rate    may reach~$5$ [$15$] new peptide bonds per second
in eukaryotes  [prokaryotes] (see~\cite{Yonath2012}).

The ribosomal density along the mRNA molecule may affect different fundamental intracellular phenomena.
A very high density can lead to ribosomal  traffic jams, collisions and
abortions.
It may also contribute  to  co-translational misfolding of proteins.
On the other hand, a very low ribosomal density
may lead to a low production rate, and a high degradation rate of mRNA molecules~\cite{Drummond2008,Kimchi-Sarfaty2013,Kurland1992,Edri2014,Zhang2009,Tuller2015,Proshkin2010}. Thus, a natural and important   question
is what  ribosomal density   optimizes one (or more) intracellular phenomena, for example,  the protein production rate.
Optimizing the protein production rate is also an important challenge in synthetic biology and biotechnology, where a
standard  objective is to maximize
 the translation efficiency and protein levels of heterologous genes in a new host (see, e.g.,~\cite[Chapter~9]{gene_cloning_book}.

In this paper, we analyze the density that maximizes the translation rate
using a mathematical model of ribosome flow along the mRNA molecule.
A standard mathematical model for ribosome flow  is the
  \emph{totally asymmetric simple exclusion process}~(TASEP) \cite{Shaw2003,TASEP_tutorial_2011}.
 In this model,  particles   hop unidirectionally
 along an ordered lattice of~$L$ sites. Every site can be either free or occupied by a particle, and a particle can only
   hop to a free site. This  simple exclusion principle
   models particles that have ``volume'' and thus cannot overtake one  other.
   The hops are stochastic, and the rate of hoping from site~$i$ to site~$i+1$  is denoted by~$\gamma_i$.
   A particle can hop to [from] the first [last] site of the lattice at a rate~$\alpha$   [$\beta$].
   The average flow through the lattice converges to a steady-state value that depends on
	the parameters~$L,\alpha, \gamma_1,\dots,\gamma_{L-1},\beta$.
	Analysis of TASEP in non trivial, and closed-form results have been obtained mainly for the homogeneous~TASEP (HTASEP), i.e.
	for the case where all the~$\gamma_i$s are assumed to be equal.
	
  TASEP has become a
 fundamental model in non-equilibrium statistical mechanics, and has been
 applied to model numerous natural and artificial  processes~\cite{TASEP_book}.
In the context of translation, the lattice  models the mRNA molecule, the particles are ribosomes,
   and simple exclusion means that a ribosome  cannot overtake a ribosome in front of it.

TASEP has two standard  
 configurations. In TASEP with \emph{open boundary conditions} the two sides of the chain are connected to two particle reservoirs, and particles can hop into the chain (if the first site is empty) and out of the chain (if the last site is full).  In TASEP with \emph{periodic boundary conditions} the chain is closed, and a particle that hops from the last site returns to the first one. Thus, here the particles hop around a ring, and the number of particles is conserved.

\begin{figure*}[t]
\centering
 \scalebox{0.85}{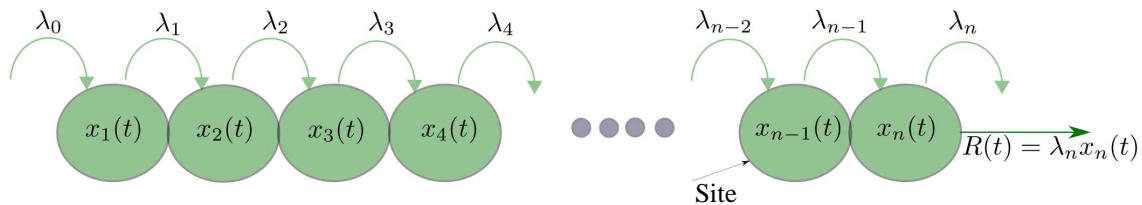}
%\scalebox{0.8}{\input{rfm_sys1_bw.eps_tex}}
\caption{The RFM models unidirectional flow along a chain of $n$   sites.
The state variable~$x_i(t)\in[0,1]$ represents
 the density of site $i$ at time $t$. The parameter~$\lambda_i>0$ controls the transition   rate  from  site~$i$ to site~$i+1$, with~$\lambda_0$ [$\lambda_n$] controlling    the initiation [exit] rate. The output rate at time $t$ is $R(t) =\lambda_n x_n(t)$.}\label{fig:rfm}
\end{figure*}

The \emph{ribosome flow model}~(RFM)~\cite{reuveni} is a continuous-time, deterministic,
 compartmental model for the unidirectional  flow of ``material" along an open chain of $n$ consecutive compartments (or sites).
The~RFM can be derived via  a dynamic
 mean-field approximation of~TASEP with open boundary conditions~\cite[section 4.9.7]{TASEP_book}~\cite[p. R345]{solvers_guide}.
The RFM includes~$n$ state-variables, denoted~$x_1(t),\dots x_n(t)$,
with $x_i(t) $ describing the   amount (or density)
of ``material'' in site~$i$ at time~$t$, normalized such that $x_i(t)=1$ [$x_i(t)=0$]
 indicates that site~$i$ is completely full [completely empty] at time~$t$.
In the~RFM, the two sides of the chain are connected to two particle reservoirs.
A   parameter~$\lambda_i>0$, $i=0,\dots,n$, controls the transition rate from site~$i$ to site~$i+1$,
 where~$\lambda_0$ [$\lambda_n$] is
the initiation [exit] rate (see Fig.~\ref{fig:rfm}).

In the  \emph{ribosome flow model on a ring (RFMR)}~\cite{rfmr}
 the particles exiting the last site enter the first site. This
 is the mean-field approximation of~TASEP with periodic boundary conditions. Since the number of particles is conserved, 
the~RFMR admits a first integral. Both the~RFM and~RFMR are cooperative dynamical systems~\cite{hlsmith}, but
their dynamical properties   turn out to be quite different~\cite{rfmr}.

The RFM [RFMR] has been  applied  to model and analyze  ribosome
 flow along an open [circular] mRNA molecule during   translation. Indeed,
it is well known that in eukaryotes the~mRNA is often (temporarily) circularized, for example,
 by translation initiation factors~\cite{Wells1988}. In addition, circular RNA forms
appear  in all domains of life~\cite{Danan2012,Cocquerelle1993,Cell1993,Burd2010,Hensgens1983,Bretscher1968,Bretscher1969}.

Here, we
use the~RFM  [RFMR] to analyze the ribosomal  density along a linear [circular]
mRNA molecule that
maximizes the steady-state protein production rate. We refer to this density as the \emph{optimal density}.
This problem has already been studied before.
For example, Zouridis and Hatzimanikatis~\cite{Zouridis2007717} derived
  a deterministic, sequence-specific kinetic model  for translation and studied
	  the effect of the average ribosomal density on the steady-state production rate. Their
		model assumes homogeneous elongation rates and open-boundary conditions, and includes all the elementary steps involved in the elongation cycle at every codon. Their simulations suggest
		 that  there exists a unique average density that corresponds to a maximal production rate, see Figures~2A and~5A
		in~\cite{Zouridis2007717} (see also~\cite{racle12}).
		
		The RFM and RFMR are simpler models and thus allow to rigorously prove
		several analytic results on the optimal density.
%%%
 For a circular mRNA, we prove
that there always exists a unique optimal density
 that maximizes the steady-state production rate, and that it
 can be determined efficiently using a simple ``hill climbing" algorithm. In addition, we show that  under certain symmetry conditions on the rates
the optimal density is one half of the maximal possible density.

\begin{figure}[t]
 \begin{center}
\includegraphics[width=5cm,height=5cm]{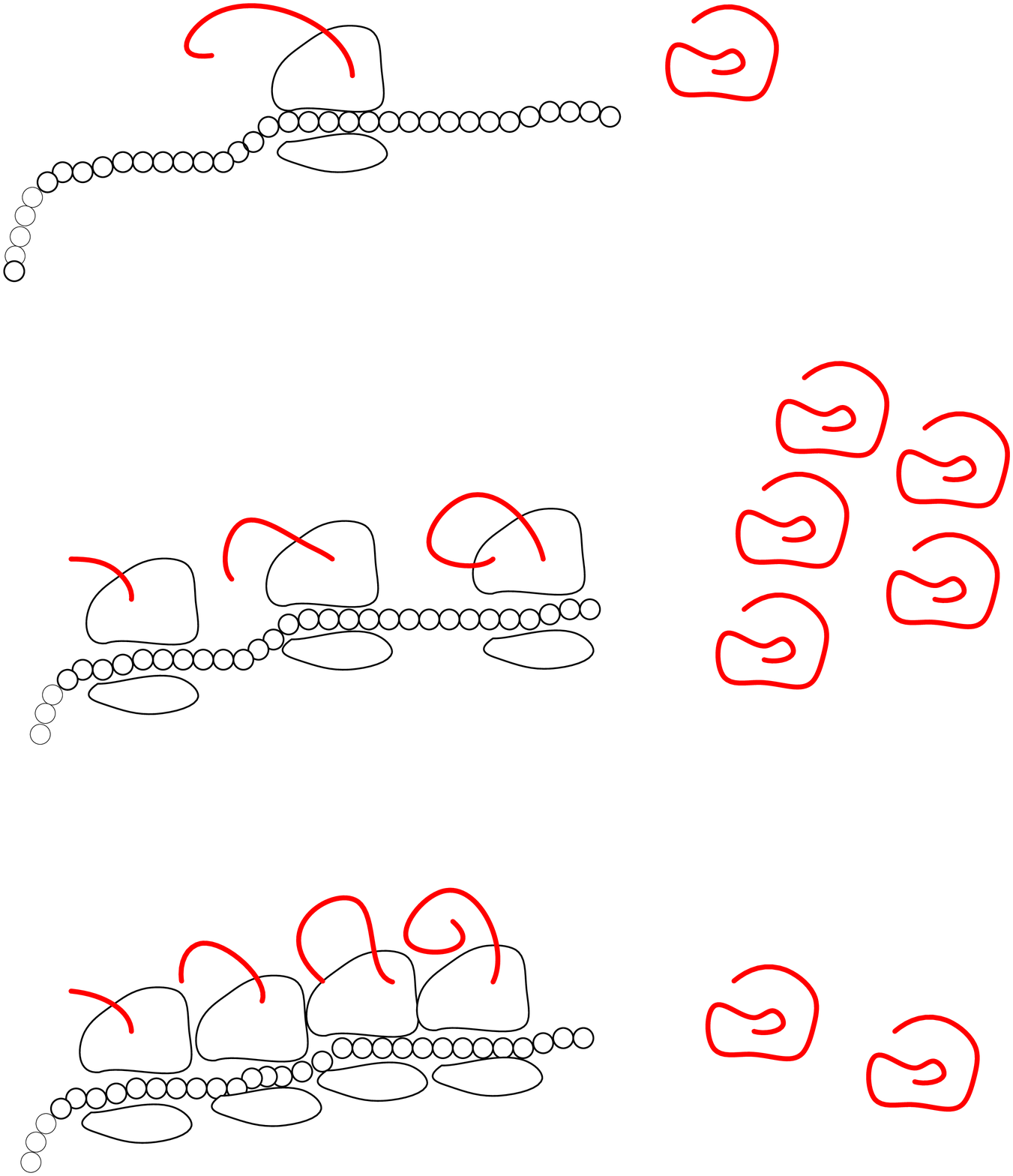}
\caption{Ribosome density and production rate. Too  few ribosomes (upper figure)
lead to a low production rate, as do  too many ribosomes (lower figure) due to traffic jams along the mRNA.
Optimal production is   achieved when the density is one half of the maximal possible density (middle figure).
 }\label{fig:illustratio}
 \end{center}
\end{figure}

 In the case of a linear mRNA molecule, we prove that when  the initiation and elongations rates
are chosen to optimize the production rate,
 under an affine constraint on the rates, the  corresponding 
 optimal density is one half of the maximal possible density
 (see  Fig.~\ref{fig:illustratio}).

The remainder of this paper  is organized as follows.
The next section briefly reviews   the~RFM and the~RFMR.
Section~\ref{sec:main} describes our main results.
 The proofs of all the results are placed in the Appendix.
The final section summarizes the results, describes their biological implications,
and suggests  several
directions for further research.

%%%%%%%%%%%%%%%%%%%%%%%%%%%%%%%%%%%
\section{The Ribosome Flow Model}\label{sec:rfm}
%%%%%%%%%%%%%%%%%%%%%%%%%%%%%%%%%%%
The dynamics of the~RFM with $n$ sites is given by $n$ nonlinear first-order ordinary differential equations:
\begin{align}\label{eq:rfm}
%%%
                    \dot{x}_1&=\lambda_0 (1-x_1) -\lambda_1 x_1(1-x_2), \nonumber \\
                    \dot{x}_2&=\lambda_{1} x_{1} (1-x_{2}) -\lambda_{2} x_{2} (1-x_3) , \nonumber \\
                    \dot{x}_3&=\lambda_{2} x_{ 2} (1-x_{3}) -\lambda_{3} x_{3} (1-x_4) , \nonumber \\
                             &\vdots \nonumber \\
                    \dot{x}_{n-1}&=\lambda_{n-2} x_{n-2} (1-x_{n-1}) -\lambda_{n-1} x_{n-1} (1-x_n), \nonumber \\
                    \dot{x}_n&=\lambda_{n-1}x_{n-1} (1-x_n) -\lambda_n x_n.
\end{align}
%%%%%%%
If we define~$x_0(t):=1$ and $x_{n+1}(t):=0$
then~\eqref{eq:rfm} can be written more succinctly as
\be\label{eq:rfm_all}
\dot{x}_i=\lambda_{i-1}x_{i-1}(1-x_i)-\lambda_i x_i(1-x_{i+1}),\quad i=1,\dots,n.
\ee
This equation  can be explained as follows.
The change in the density in site~$i$ is the flow from site~$i-1$ to site~$i$
minus the flow from site~$i$ to site~$i+1$. The latter
  is~$\lambda_{i} x_{i}(t)(1 - x_{i+1}(t) )$. This flow is proportional to $x_i(t)$, i.e. it increases
	with the density at site~$i$, and to $(1-x_{i+1}(t))$, i.e. it decreases as site~$i+1$ becomes fuller.
In particular, when the site is completely full, i.e.~$x_{i+1}(t)=1$, there is no flow
into this site.
 This corresponds to a ``soft''  version of a simple exclusion principle: the flow of
particles into a site decreases as that site becomes fuller. Note that the maximal possible
  flow  from site~$i$ to site~$i+1$  is the~$i$th transition rate~$\lambda_i$.
The output rate from the chain is  $R(t):=\lambda_n x_n(t)$.

Let~$x(t,a)$ denote the solution of~\eqref{eq:rfm}
at time~$t \ge 0$ for the initial
condition~$x(0)=a$. Since the  state-variables correspond to normalized occupation levels,
  we always assume that~$a$ belongs to the  closed $n$-dimensional
  unit cube:
$
           C^n:=\{x \in \R^n: x_i \in [0,1] , i=1,\dots,n\}.
$
It is straightforward to verify that this implies that~$x(t,a) \in C^n$ for all~$t\geq0$.
In other words,~$C^n$ is an invariant set of the dynamics~\cite{RFM_stability}.

Let~$\Int(C^n)$ denote the interior of~$C^n$.
It was shown in~\cite{RFM_stability} that the RFM is a
\emph{cooperative  dynamical system}~\cite{hlsmith}
and that this implies that~\eqref{eq:rfm}
admits a \emph{unique} steady-state point~$e =e( \lambda_0,\dots,\lambda_n )\in \Int(C^n)$
 that is globally asymptotically stable, that is,
$\lim_{t\to \infty} x(t,a)=e$ for all~$a\in C^n$ (see also~\cite{RFM_entrain}).
In particular, this means that the production rate converges to the steady-state value:
\be \label{eq:defr}
R:=\lambda_n  {e}_n.
\ee

For~$x=e$ the left-hand side of all the equations
in~\eqref{eq:rfm} is zero, so
%%%
%%%
\begin{align} \label{eq:ep}
%%%
                      \lambda_0 (1- {e}_1) & = \lambda_1 {e}_1(1- {e}_2)\nonumber \\&
                      = \lambda_2  {e}_2(1- {e}_3)   \nonumber \\ & \vdots \nonumber \\
                    &= \lambda_{n-1} {e}_{n-1} (1- {e}_n) \nonumber \\& =\lambda_n  {e}_n \nonumber \\&=R.
%%%
%%%
\end{align}
This yields
\begin{align}\label{eq:list}
                             {e}_n & = R/\lambda_n, \nonumber  \\
                             {e}_{n-1} & = R / (\lambda_{n-1} (1- {e}_n) ),  \nonumber\\
                            & \vdots \nonumber \\
                             {e}_{2} & = R / (\lambda_{2} (1- {e}_3) ), \nonumber\\
                             {e}_{1} & = R / (\lambda_{1} (1- {e}_2) ),
%%
%%%
\end{align}
and \be \label{eq:also}
                             {e}_1= 1-R/ \lambda_0 .
\ee

Combining~\eqref{eq:list} and~\eqref{eq:also} provides an elegant
\emph{finite continued fraction}~\cite{waad} expression for~$R$:

\begin{align} \label{eq:cf}
                0&= 1-\cfrac{R/ \lambda_0 }
                                  {  1-\cfrac{R / \lambda_1}
                                  {1-\cfrac{R / \lambda_2}{\hphantom{aaaaaaa} \ddots
                             \genfrac{}{}{0pt}{0}{}
                             {1-\cfrac{R/\lambda_{n-1}}{1-R/ \lambda_n.}} }}}
\end{align}
Note that this equation admits several solutions for~$R$, however, we are interested only in the unique feasible solution, i.e. the solution corresponding to~$e \in \Int(C^n)$.
Note also that~\eqref{eq:cf} implies that
 \be\label{eq:rishomog}
R(c \lambda_0,\dots,c \lambda_n)= c R (  \lambda_0,\dots,  \lambda_n)  ,\quad \text{for all }c>0,
 \ee
that is,~$R(\LMD)$ is a \emph{homogeneous  function} of degree one.  Ref.~\cite{rfm_max} proved that $R(\LMD)$ is a \emph{strictly concave} function on $\R^{n+1}_{++}$.

\subsection{Ribosome Flow Model on a Ring}
%%%%%%%%%%%%%%%%%%%%%%%%%%%%%%%%%%%
If we consider the RFM with the additional assumption that all the ribosomes leaving site~$n$
circulate back to site~$1$ then we obtain the RFMR:
%%%
\begin{align}\label{eq:rfmr}
%%%
                    \dot{x}_1&=\lambda_n x_n (1-x_1) -\lambda_1 x_1(1-x_2), \nonumber \\
                    \dot{x}_2&=\lambda_{1} x_{1} (1-x_{2}) -\lambda_{2} x_{2} (1-x_3) , \nonumber \\
                             &\vdots \nonumber \\
                    \dot{x}_n&=\lambda_{n-1}x_{n-1} (1-x_n) -\lambda_n x_n (1-x_1) .
%%%
%%%
\end{align}

This can also be written succinctly  as~\eqref{eq:rfm_all}, but now  with every index   interpreted modulo~$n$. In particular, $\lambda_0$
[$x_0$] is replaced by~$\lambda_n$ [$x_n$].

For~$p\in\R$, let~$p_n$ denote the column vector~$\begin{bmatrix} p&p&\dots&p \end{bmatrix}^T\in\R^n$.
 Eq.~\eqref{eq:rfmr} implies that
\[
           \frac{d}{dt} ( 1_n^T   x(t) )\equiv 0, \text{ for all }t\geq 0,
\]
so  the \emph{ribosome density}~$H(x):=1_n^T x$  is conserved, i.e.
\be\label{eq:conser}
 H(x(t)) = H(x(0)),\quad \text{for all } t\geq 0.
\ee
The dynamics of the RFMR thus redistributes the particles between the sites,
but without changing  ribosome density.
 In the context of translation, this means that
the total number of ribosomes on the  (circular) mRNA is conserved.

For~$s\in[0,n]$, denote the \emph{$s$ level set} of~$H$ by
\[
L_s:=\{  y \in C^n: 1_n^T y= s   \}.  
\]
It was shown in~\cite{rfmr} that the RFMR is a strongly cooperative
 dynamical system, that every level set~$L_s$  contains a unique equilibrium  point~$e=e(s,\lambda_1,\dots,\lambda_n)$, and that any trajectory of the RFMR  emanating  from any~$x(0)\in L_s$ converges to this   equilibrium point.
For example if~$s=0$, corresponding to the initial condition~$x(0)=0_n$,
then~$x(t)\equiv 0_n$ for all~$t\geq 0$, so~$e=0_n$. Similarly,~$s=n$ corresponds to the
initial condition~$x(0)=1_n$ and then clearly~$x(t)\equiv 1_n$ for all~$t\geq0$, so~$e=1_n$.
Since these two cases are trivial, below we will always assume that~$s\in(0,n)$.
In this case,~$e \in \Int(C^n)$.

%% The RFMR, just like the RFM, entrains to a periodic excitation of its rates~\cite{rfmr}.

 Let~$R=R(s,\lambda_1,\dots,\lambda_n)$ denote the steady-state production rate in the RFMR for~$x(0) \in L_s$.
It is straightforward to verify that for any~$c>0$
\be\label{eq:rfmrscale}
R(s,c \lambda_1,\dots, c \lambda_n)=
cR(s,\lambda_1,\dots,\lambda_n)  .
\ee

For more on the analysis of the RFM and the RFMR using tools from systems and control theory, see~\cite{zarai_infi,rfm_max,RFM_sense,rfmr,RFM_feedback,rfm_control}.
For a general discussion on using systems and control theory in systems biology see the excellent survey
 papers by Sontag~\cite{sontag_sys_bio,SONTAG2005396}.

The RFM models translation on a single isolated mRNA molecule. A network of~RFMs, interconnected through
 a common pool of ``free'' ribosomes has been used to model
simultaneous translation of several   mRNA molecules while competing for the available ribosomes~\cite{RFM_model_compete_J}.
It is important to note that many analysis results for the RFM, RFMR, and networks of RFMs hold
for any set of transition rates. 
This is in contrast to the analysis results on the TASEP model. 
Rigorous analysis of TASEP seems to be tractable only under the assumption
  that the internal hopping rates are all equal (i.e. the homogeneous case).

The next section describes our main results on the optimal ribosome density.

%%%%%%%%%%%%%%%%%%%%%%%%%%%%%%%%%%%
\section{Main Results}\label{sec:main}
%%%%%%%%%%%%%%%%%%%%%%%%%%%%%%%%%%%
Let
$\rho(t):=\frac{1}{n} (1_n^T x(t))$ denote the average ribosome density along the mRNA molecule
at time~$t$.
Recall that for every set of  parameters in  our models the state-variables converge
to a steady-state~$e$. In particular,~$\rho(t)$ converges to the
steady-state average  ribosomal density:
\[
\rho:=\frac{1}{n} (1_n^T e ).
\]
Note that since~$e_i \in[0,1]$ for all~$i$, $\rho \in [0,1]$.
We are interested in analyzing the density that is obtained when the parameter values in the model
are the ones that
  maximize the steady-state
production rate.

%%%%%%%%%%%%%%%%%%%%%%%%%%%%%%%%%%%
\subsection{Optimal   Density in the RFMR}
%%%%%%%%%%%%%%%%%%%%%%%%%%%%%%%%%%%
Recall that in the RFMR  the dynamical behavior depends on the rates and the
 quantity~$s:=1_n^T x(0)$. The ribosomal
 density is constant:~$\rho(t)\equiv s/n$. Fix arbitrary transition rates~$  \lambda_i>0$, $i=1,\dots,n$, and let
 $R(s):=R(s;\LMDRFMR)$ and $e(s):=e(s;\LMDRFMR)$ denote the steady-state production rate and the ribosomal densities, respectively, as a function of~$s$.
The next result shows that
there always exists a unique density~$\rho^*=s^*/n$ that corresponds to a maximal steady-state production rate.

%The next result shows that
%  there is always a unique density that yields a  maximal
% steady-state production rate.

\begin{Proposition}\label{prop:rfmr_dense}
%%%%%%%%%%%%
For any set of rates~$\lambda_i>0$ in the RFMR there exists a unique value~$s^*=s^*(\lambda_1,\dots,\lambda_n)\in(0,n) $ that maximizes~$R(s)$. Furthermore, for this optimal value~$e^*:=e(s^*)$ and~$R^*:=R(s^*) $ satisfy
\be\label{eq:prode}
								   e_1^*\dots e_n^* =(1-e_1^*) \dots (1-e_n^*)  ,
\ee
  and
\be\label{eq:rnstar}
(R^*)^n=(\lambda_1\dots\lambda_n) (e_1^*\dots e_n^*)^2.
\ee
\end{Proposition}
The proof of this result (given in the Appendix) shows that~$R(s)$ is strictly increasing on~$[0,s^*)$
and strictly decreasing on~$(s^*,n]$,
 so  a   simple ``hill climbing'' algorithm
can be used to find~$s^*$.
%%%%

The optimality condition~\eqref{eq:prode} can be explained as follows. If~$s$ is very small
then there will not be enough ribosomes on the circular mRNA and the production rate will be small (for example, for~$s=0$ we have~$e=0_n$, and
thus~$R=\lambda_1 e_1(1-e_2)=0$). In this case, the product of the~$e_i$s
is   small, so~$ e_1 \dots e_n   < (1-e_1 ) \dots (1-e_n )$ and~\eqref{eq:prode} does not hold. If~$s$ is very large
  traffic jams evolve on the mRNA  and again the production rate will be small (for example, for~$s=n$ we have~$e=1_n$, and
thus~$R=\lambda_1 e_1(1-e_2)=0$). In this case,~$ e_1 \dots e_n   > (1-e_1 ) \dots (1-e_n )$ and~\eqref{eq:prode} does not hold. Thus,~\eqref{eq:prode}
describes the   point where the balance between too few and too many ribosomes is optimal.

\begin{figure}[t]
 \begin{center}
\includegraphics[width= 7cm,height=6cm]{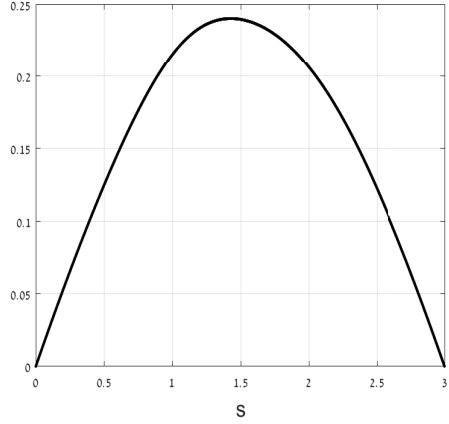}
\caption{Steady-state production rate~$R(s)$ as a function of~$s$ for the RFMR in Example~\ref{exa:strcus}. }\label{fig:rfmr_maxR_n5}
 \end{center}
\end{figure}

\begin{figure}[t]
\centering
 \scalebox{1.5}{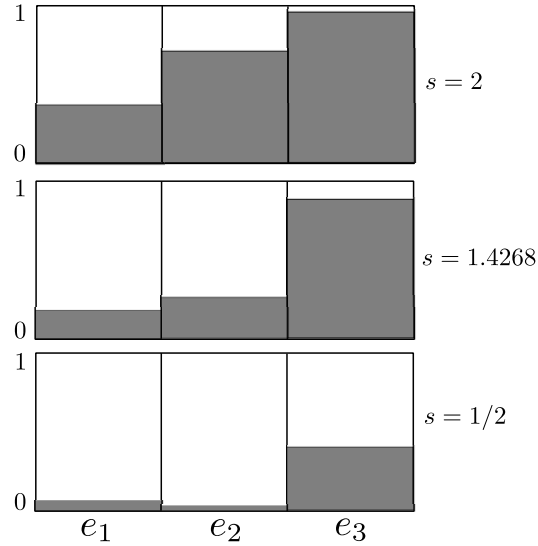}
\caption{Equilibrium point~$e$ in Example~\ref{exa:strcus} for three different~$s$ values.}\label{fig:opt_e_n3}
\end{figure}

The next example demonstrates   Proposition~\ref{prop:rfmr_dense} in a special case.
%%%%%%%
\begin{Example}\label{exa:scase}
%%%
Consider an RFMR with~$\lambda_1=\dots=\lambda_n $, i.e. all the rates are equal. Denote their common value by~$\lambda_c$.
Then it follows from~\eqref{eq:rfmr} that~$1_n c$, $c>0$, is an   equilibrium point. By uniqueness of the equilibrium point in every level set of~$H$
this implies that~$e=(s/n) 1_n$, and thus~$R=\lambda_n e_n(1-e_1)=\lambda_c (s/n)(1-(s/n))$. Thus,
$
		\frac{\partial R}{\partial s}=\frac{\lambda_c}{n^2}(n-2s)$, so
 $R(s)$ is strictly increasing [decreasing] on~$s\in[0,n/2]$ [$s\in[n/2,n]$]
and therefore
 attains a unique maximum at~$s^*=n/2$. Then~$e^*:=e(s^*)=(1/2)1_n$ and~$R^*:=R(s^*)=\lambda_c/4$, and it is straightforward to verify that
\eqref{eq:prode} and \eqref{eq:rnstar}
hold. Note also that  $\frac{\partial^2 R}{\partial s^2}=-\frac{2\lambda_c}{n^2}<0$, implying that~$R(s)$ is
a strictly concave function.\hfill{$\square$}
%%%%
\end{Example}
The next example demonstrates the dependence of~$R(s)$ on~$s$ when the rates
 are not homogeneous.
%%%%%%%%%%%%%%%%%%%%%%%%%%%%%%%
\begin{Example}\label{exa:strcus}
Consider an  RFMR  with dimension $n=3$ and transition
 rates~$  \lambda_1=2$, $  \lambda_2=6$, and $  \lambda_3=1/3$. Fig.~\ref{fig:rfmr_maxR_n5}
 depicts~$R(s)$ for~$s\in[0,3]$. It may be seen
that~$R(s)$ attains a unique maximum at~$s^* =1.4268 $ (all numerical values in this paper are to four digit accuracy).
The corresponding equilibrium point is~$e^*=\begin{bmatrix}  0.1862   & 0.3539&    0.8867    \end{bmatrix}^T$,
and the optimal production rate is~$R^*=\lambda_1 e^*_1(1-e^*_2) = 0.2405$.
Fig.~\ref{fig:opt_e_n3} depicts a histogram of the equilibrium point~$e$ for three values of the level set parameter: $s=1/2$, $s=1.4268$,
and~$s=2$.  Note that~$e_3$ is the maximal entry in~$e$ for all~$s$. This is
due to fact that the entry rate~$\lambda_2=6$ into site~$3$ is high, and the
exit rate~$\lambda_3=1/3$ from site~$3$   is low.~\hfill{$\square$}
\end{Example}

In order to better understand Fig.~\ref{fig:opt_e_n3} note that
the equilibrium point in the~RFMR
 satisfies
\[
e_1+\dots+e_n=s,
\]
and, by~\eqref{eq:rfmr},
\begin{align}\label{eq:rfmr_e}
%%%
                     \lambda_n e_n (1-e_1) &=\lambda_1 e_1(1-e_2), \nonumber \\
                                           &=\lambda_{2} e_{2} (1-e_3) , \nonumber \\
                             &\vdots \nonumber \\
                     &=\lambda_{n-1}e_{n-1} (1-e_n)  .
%%%
%%%
\end{align}
Let~$k_i:=   \lambda_1\dots\lambda_{i-1} \lambda_{i+1}\dots\lambda_n $, $i=1,\dots,n$,
and let~$\mu:=\sum_{i=1}^n k_i$.
%%%%%
If~$s\approx 0$ then all the~$e_i$s will be small, so we can ignore the terms~$1-e_i$ 
in~\eqref{eq:rfmr_e}, and
this yields~$e_i\approx \frac{k_i s}{\mu}$, $i=1,\dots,n$.
A similar argument shows that  if~$s\approx n$ 
then~$e_i \approx 1-\frac{k_{i-1} (n-s)}{\mu}$, $i=1,\dots,n$. 
 For the particular case in Example~\ref{exa:strcus} this implies that
when~$s\approx 0$ $e\approx (s/22)\begin{bmatrix}  3&1&18  \end{bmatrix}^T$.
In particular,~$e_2<e_1<e_3$. When~$s\approx 3$, $e\approx (s/22)\begin{bmatrix}  18s-32& 3s+13&s+19  \end{bmatrix}^T$.
In particular,~$e_1<e_2<e_3$.

For small values of~$n$ it is possible to give more explicit results.
\begin{Fact}\label{fact:sim}
%%%%%%%
For an RFMR with~$n=2$ the optimal values are~$s^*=1$ and
\be\label{eq:rstar2}
R^*=  \frac{   \lambda_1 \lambda_2}{  (\sqrt{\lambda_1}+\sqrt{\lambda_2})^2}       .
\ee
%%%%
For an RFMR with~$n=3$ the optimal production rate satisfies:
\be\label{eq:rstar3}
     2\sqrt{\lambda_1\lambda_2\lambda_3} (R^*)^{3/2}+(\lambda_1
\lambda_2+\lambda_1\lambda_3+\lambda_2\lambda_3)R^*-\lambda_1\lambda_2\lambda_3 =0.
\ee
%%%%%%
%%%%
\end{Fact}

Let~$e_i':=\frac{\partial  }{\partial s}e_i$ denote the sensitivity of~$e_i$ with respect to a change in the total density~$s$.
The next results provides an expression for these sensitivities at the equilibrium point
corresponding to the optimal density.
\begin{Proposition}\label{prop:rfmr_sense}
Consider an RFMR with dimension~$n$. Fix rates~$\lambda_i>0$, and let~$s^*=s^*(\lambda_1,\dots,\lambda_n)$
and~$e^*=e^*(\lambda_1,\dots,\lambda_n)$ be as defined in Proposition~\ref{prop:rfmr_dense}.
Then~$(e^*)'=\frac{v}{1_n^T v}$, where
\be\label{eq:vv}
v:=\begin{bmatrix}   \frac{e_1^*  \dots e^*_{n-1}}{(1-e_2^*)\dots(1-e^*_n)}&
\frac{e_2^*  \dots e^*_{n-1}}{(1-e_3^*)\dots(1-e^*_n)}&\dots&
\frac{  e^*_{n-1}}{  1-e^*_n }&
1
   \end{bmatrix}^T.
\ee
\end{Proposition}

\begin{Example}\label{exa:stam}
%%%%%%%%%%%%%%%%%%%%%%%%%%%
Consider again the   RFMR  in Example~\ref{exa:strcus}.
Recall that here~$s^* =1.4268 $ and~$e^*=\begin{bmatrix}  0.1862   & 0.3539&    0.8867    \end{bmatrix}^T$.
Substituting this in~\eqref{eq:vv}  yields
$
v=\begin{bmatrix} 0.9002& 3.1236& 1 \end{bmatrix}^T,
$
so~$(e^*)'=\begin{bmatrix} 0.1792& 0.6218& 0.1991 \end{bmatrix}^T$.
This means that if we change the density from~$s^*$ to~$\bar s:=s^*+\varepsilon$ then the steady-state production rate
  changes from~$R^*$ to
\begin{align*}
\bar R&= \lambda_1 \bar e_1 (1-\bar e_2)\\
    &	=\lambda_1 (e_1^*+\varepsilon (e_1^*)')(1-   e_2^* -  \varepsilon (e_2^*)') +O(\varepsilon^2)\\
    &	= R^*+ \lambda_1\varepsilon (  (1-e^*_2)(e_1^*)'-e_1^*(e_2^*)'      )  +O(\varepsilon^2),
\end{align*}
and substituting the numerical values yields
\[
			\bar R=R^*+O(\varepsilon^2).
\]
%%%
Indeed, this agrees with the fact that   the graph of~$R(s)$
attains a maximum at~$s^*$. \hfill{$\square$}
\end{Example}

In  Example~\ref{exa:strcus} above   the optimal value~$s^*$ is close, but not equal to~$n/2=3/2$.
The next result provides a symmetry condition   guaranteeing
 that~$s^*=n/2$, that is,   that the
optimal density is equal to one half of the maximal possible density.
\begin{Proposition}\label{prop:rfmr_half}
If the transition rates in the~RFMR  
satisfy
\be\label{eq:condf}
\lambda_i=\lambda_{n-i},  \quad i=1,\dots,n,
\ee
 then~$s^*=n/2$ and~$e^*_i=e^*_{n+1-i}$ for all~$i$.
%%%
\end{Proposition}
Thus, in this case the optimal mean density is~$\rho^*=(n/2)/n=1/2$. Note that condition~\eqref{eq:condf} always holds for~$n=2$.
Also, since a cyclic permutation of the rates leads to an RFMR with the same behavior,
it is enough that~\eqref{eq:condf} holds for some cyclic permutation of the rates.
 For~$n=3$ this holds if at least  two of the rates~$\lambda_1,\lambda_2,\lambda_3$
are equal.

We note that a  result  similar to Proposition~\ref{prop:rfmr_half} is known for the
  \emph{homogeneous}
 TASEP with periodic boundary conditions, i.e. that a loading of~$50\%$ maximizes the steady-state flow
 (see, for example, the fundamental diagram in~\cite[Figure~4.1]{TASEP_book}).

%%%%%%%%%%%%%%%%%%%%%%%%%%%%%%%%%%%
\subsection{Optimal  Density in the RFM}
%%%%%%%%%%%%%%%%%%%%%%%%%%%%%%%%%%%
Due to the open boundary conditions in the RFM, the number of particles along the chain is not conserved.
Thus, in this section we analyze the steady-state densities corresponding to the rates that yield a  maximal steady-state production
rate.
To do this, we recall the optimization problem posed in~\cite{rfm_max}.
%%%%%%
 \begin{Problem}\label{prob:max}
 Fix  parameters $b, w_0,w_1,\dots,w_n  >0$.
Maximize~$R =R(\LMD)$, with respect to its parameters~$\LMD$, subject to the constraints:
\begin{align}\label{eq:constraint}
\sum_{i=0}^n w_i\lambda_i  & \leq  b, \\
\LMD &\geq 0.\nonumber
\end{align}
\end{Problem}
%%%%%%%%
In other words, maximize the steady-state production rate   given an affine constraint on the rates.
Here~$b$ is the ``total biocellular budget'', and the positive values $w_i$, $i=0,\dots,n$, can be used to provide a
different weighting to the different rates.

This formulation is motivated   by the fact that the biological resources are of course limited.
 For example, all~tRNA molecules are transcripted by the same transcription factors~(TFIIIB) and by~RNA polymerase III. Hence, if the production of a specific~tRNA is increased then the production of some other~tRNA must  decrease.  The total cost~$b$ captures this, as any increase in one of the~$\lambda_i$s must
be compensated by a decrease in some other rate.
%%%%

Problem~\ref{prob:max} formalizes, using the RFM, an
 important problem in both systems biology and biotechnology, namely,  determine the transition rates
that maximize
   the protein production rate, given the limited
biomolecular budget.

It has been shown in~\cite{rfm_max} that the optimal solution $\LMDSTAR$ always satisfies~$\sum_{i=0}^n w_i\lambda_i^*= b$.
 Of course, by scaling the~$w_i$s we may always assume that~$b=1$. Combining this with the strict concavity of the steady-state production rate~$R(\LMD)$ in the~RFM 
implies that
 Problem~\ref{prob:max} is a \emph{convex optimization problem} that admits a unique optimal solution $\lambda^*\in\R^{n+1}_{++}$. This solution can thus be found efficiently using numerical algorithms that scale well with $n$.
Here, our goal is to determine what is the steady-state density when the optimal rates are used, that is, when the rates are the solution of Problem~\ref{prob:max}. We refer to this as the optimal density.
 Let $e^*_i$, $i=1,\dots,n$, denote the steady-state density at site $i$ corresponding to the optimal rates $\LMDSTAR$.

\begin{figure}[t]
 \begin{center}
\includegraphics[width= 8cm,height=7cm]{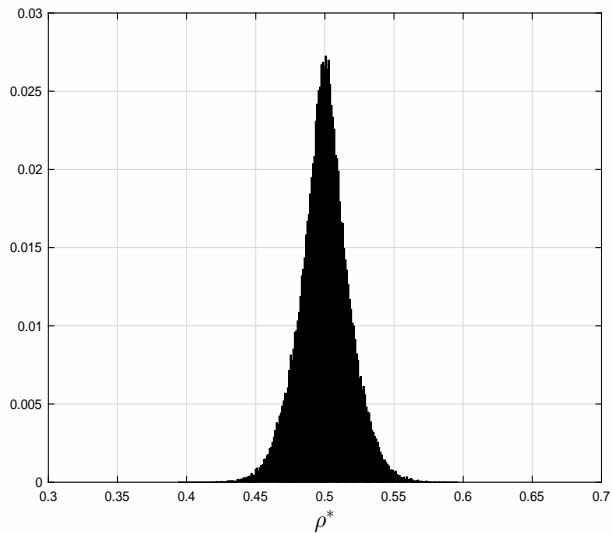}
\caption{Normalized histogram of the value~$\rho^*$ in Example~\ref{exa:histo}.}\label{fig:rfm_n11_wu}
 \end{center}
\end{figure}

\begin{Example}\label{exa:histo}
%%%%%%%%%%%%%%%
 Using a simple numerical algorithm  we solved~$10^5$ instances of Problem~\ref{prob:max} for an RFM with length~$n=11$ and total budget~$b=1$.
In each instance
  the weights~$w_i$ were drawn independently from a
 uniform distribution  over the interval~$[0,1]$. For each instance, we computed the optimal rates~$\lambda^*_i$s
and the corresponding mean steady-state optimal density~$\rho^*:=\frac{1}{n} \sum_{i=1}^n e_i^*$.
 Fig.~\ref{fig:rfm_n11_wu} depicts a normalized histogram (that is, the empirical probability)
of   the~$10^5$ values of~$\rho^*$.
  It may be observed that typically~$\rho^*$ is close to~$1/2$.
Similar results are obtained when the weights are drawn using other statistics, e.g. exponential,
Rayleigh, and Gamma  distributions. \hfill{$\square$}
\end{Example}

In the case where all the weights are equal we can also derive theoretical results on the structure of~$e^*$ and thus of~$\rho^*$.

\subsubsection{Homogeneous Affine Constraint}
%%%%%%%%%%%%%%%%%%%%%%%%%%%%%%
Consider the case where all the weights~$w_i$ in Problem~\ref{prob:max} are equal.
We refer to this  as the \emph{homogeneous constraint} case.
 Indeed, in this case the weights
give equal preference to all the rates, so
if the corresponding
optimal solution  satisfies~$\lambda_i^* > \lambda_j^*$ for some~$i,j$ then this implies
that, in the context of maximizing~$R$,~$\lambda_i$ is ``more important'' than~$\lambda_j$.
By~\eqref{eq:rishomog}, we may assume in this case, without loss of generality, that~$w_0=\dots=w_n=b=1$, so the 
constraint is
\be \label{eq:homog_comnst}
\sum_{i=0}^n \lambda_i \leq 1 .
\ee

\begin{Proposition}\label{prop:lambda_ratio_hom}
%%%%%%%%%%%%%%%%%%%%%%%%%%%%
Consider Problem~\ref{prob:max} with the
 homogeneous constraint~\eqref{eq:homog_comnst}. Then the optimal steady-state occupancies satisfy
\be\label{eq:eismono}
e_i^*=1-e_{n-i+1}^*,\quad i=1,\dots,n.
\ee
If $n$ is even then
\be\label{eq:eisneven}
          e_1^*> \dots >e_{\frac{n}{2}}^*>\frac{1}{2}>  e_{\frac{n}{2}+1}^*> \dots>  e_{n}^*,
\ee
and if $n$ is odd then
\be\label{eq:eisnodd}
          e_1^*> \dots > e_{\frac{n-1}{2}}^*  > e_{\frac{n+1}{2}}^*=\frac{1}{2}>    e_{\frac{n+2}{2}}^*> \dots>  e_{n}^*.
\ee
In both cases, the corresponding optimal density is~$\rho^*=1/2$.
\end{Proposition}

Proposition~\ref{prop:lambda_ratio_hom} implies that under the homogeneous
constraint  the steady-state occupancies corresponding to the optimal solution  דארןבאךט
 decrease along the chain, and are anti-symmetric with respect to the center of the chain, i.e. $e_i^*-1/2=1/2-e^*_{n-i+1}$, $i=1,\dots,n$. This immediately
implies that $\rho^*= \frac{1}{n}\sum_{i=1}^n e_i^*=1/2$.

\begin{figure}[t]
  \begin{center}
  \includegraphics[width= 7cm,height=6cm]{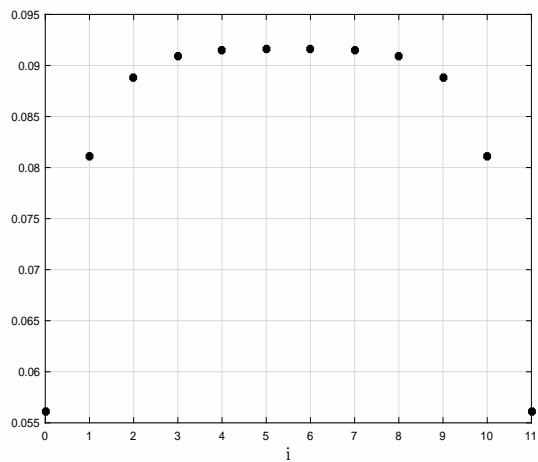}
  \caption{  Optimal rates~$\lambda_i^*$ as a function of~$i$
  for an RFM with $n=11$ and the homogeneous constraint~\eqref{eq:homog_comnst}.}\label{fig:rfm_n11_l}
  \end{center}
\end{figure}

\begin{figure}[t]
  \begin{center}
  \includegraphics[width= 7cm,height=6cm]{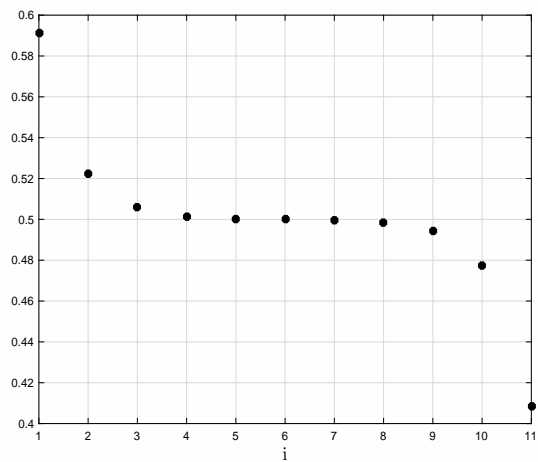}
  \caption{Optimal steady-state ribosome distribution~$e_i^*$ as a function of~$i$
  for an RFM with $n=11$ and the homogeneous constraint~\eqref{eq:homog_comnst}.}\label{fig:rfm_n11_e}
  \end{center}
\end{figure}

\begin{Example}\label{exp:rfmn11}
Consider Problem~\ref{prob:max}  for an RFM with $n=11$ and the homogeneous constraint~\eqref{eq:homog_comnst}. Fig.~\ref{fig:rfm_n11_l} depicts the optimal values~$\lambda_i^*$, $i=0,\dots,11$.
It may be seen that the~$\lambda_i^*$s are symmetric, i.e. $\lambda_i^*=\lambda_{11-i}^*$, and that they increase towards the center of the chain.
The corresponding steady-state distribution is~$e^*=
[ 0.5913$, $0.5224$, $0.5059$, $0.5016$, $0.5004$, $0.5000$,  $0.4996$, $0.4984$, $0.4941$, $0.4776$, $0.4087 ]^T$
(see Fig.~\ref{fig:rfm_n11_e}).
It may be seen that  the steady-state densities strictly decrease along the chain and are anti-symmetric with respect to the center of the chain.\hfill{$\square$}
\end{Example}

\begin{comment}
%%%%%%%%%%%
It is interesting to apply
 the analytical results obtained for the RFM to the stochastic TASEP model with open boundary conditions.
Indeed, the RFM is a mean-field approximation of TASEP with open boundary conditions, and a natural question is whether the analytical results of the RFM can
 be used as estimates to the time averaged TASEP performances. For example, do the optimal rates yield a  similar time averaged densities to that of the RFM.

Consider Example~\ref{exp:rfmn11}. We simulated TASEP with open boundary conditions, in parallel update mode, with $L=11$ sites,  and with transition probabilities $\alpha:=\lambda_0^*/\delta$, $\gamma_i:=\lambda_i^*/\delta$, $i=1,\dots,10$, $\beta:=\lambda_{11}^*/\delta$, where $\delta:=\sum_{i=0}^n \lambda_i^*$. Note that it follows from~\eqref{eq:ep} that normalizing all the $\lambda_i$s by a positive constant does not affect the steady-state densities in the RFM (i.e., using $\alpha, \gamma_1,\dots,\gamma_{10},\beta$ instead of $\LMDSTAR$ in the RFM yield that same steady-state densities). The time average of the densities obtained by simulation in TASEP, denoted by $\tau$, is:
\[
\tau=\begin{bmatrix} 0.5788, 0.5233, 0.5100, 0.5051, 0.5038, 0.4997, 0.4958, 0.4924, 0.4862, 0.4733, 0.4190 \end{bmatrix}^T,
\]
an thus $\frac{\sum_{i=1}^{11} \tau_i}{11}=0.4989$. Thus, using the (normalized) optimal rates in TASEP with open boundary conditions yield similar averaged occupancy to that of the RFM.

\end{comment}

 %%%%%%%%%%%%%%%%%%%%%%%%%%%%%%%%%%%
 %%%%%%%%%%%%%%%%%%%%%%%%%%%%%%%%%%%

Since the RFM [RFMR] is the dynamic  mean-field approximation of TASEP with open [periodic] boundary conditions,
our results  naturally lead  to questions on the optimal density in TASEP. These questions seem to be difficult to analyze rigorously. 
  We used a simple grid search to address the
problem of maximizing the steady-state flow in HTASEP (with all internal rates equal to  one)  with respect to the parameters~$\alpha$ and~$\beta$
subject to the constraint~$w_1 \alpha+w_2 \beta =b$.
For~$L=11$ and~$w_1=w_2=b=1$ the solution is~$\alpha^*=\beta^*=1/2$, and the corresponding steady-state
 occupancies (computed using~\cite[Eq. (3.65)]{solvers_guide}) are all equal to $1/2$.
%$
%\begin{bmatrix}
%0.4119 ,   0.5  ,  0.5  ,  0.5  ,  0.5, 0.5  , 0.5 ,   0.5  ,  0.5 ,   0.5   , 0.5
%\end{bmatrix}.
%$
Thus the average optimal density is~$\rho^*=1/2$.
 % and the corresponding averaged occupancies are:
%$
%\begin{bmatrix}
%0.4975 ,   0.5005  ,  0.5005  ,  0.5015  ,  0.5025  ,
% 0.5007  , 0.5000 ,   0.5004  ,  0.5006 ,   0.5002   , 0.4969
%\end{bmatrix},
%$
%and thus the average optimal density is~$\rho^*=0.5001$.

\begin{figure}[t]
  \begin{center}
  \includegraphics[width=7cm,height=6cm]{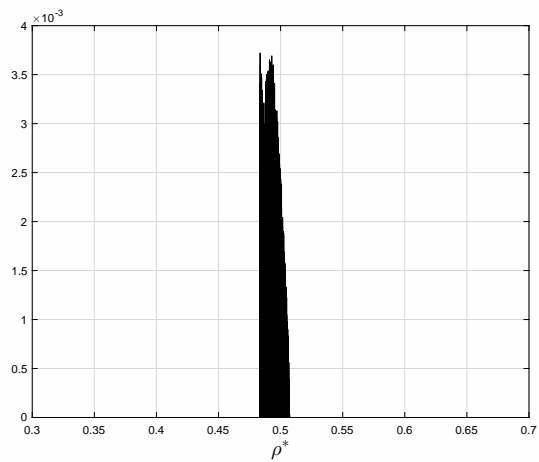}
  \caption{Normalized histogram of   steady-state  mean  
	optimal ribosome density in~HTASEP  with $N=30$ and optimal parameters.}\label{fig:htasep_N30}
  \end{center}
\end{figure}

We also ran 10000 tests with~$w_1$ and~$w_2$  chosen from
an independent  uniform distribution on~$[0,1]$.
In each case,
a simple grid-search was used to find the
 optimal rates.
Fig.~\ref{fig:htasep_N30} depicts a normalized histogram of the optimal steady-state sum of ribosome densities in an HTASEP with~$L=30$.
It may be seen that the typical optimal density is about~$1/2$. A similar result has been reported in~\cite{Marshall2014}  that used
TASSEP with
a
superposition of open and periodic boundary conditions.

These simulation
results corroborate the analytic results derived above for the~RFM and~RFMR.

 %%%%%%%%%%%%%%%%%%%%%%%%%%%%%%%%%%%
\section{Discussion}
%%%%%%%%%%%%%%%%%%%%%%%%%%%%%%%%%%%
A natural analogy  for  the cell is that of a  factory operating
 complex and inter-dependent biosynthesis
assembly processes~\cite{Pugatch24022015}. Increasing the production rate can be done by both
operating several identical processes in parallel, and by pipelining every single
 process. In the context of translation,
  many mRNA copies of the same  gene are translated in parallel, and
the same transcript is simultaneously translated by several ribosomes.
A natural question is what is the
density of ribosomes along the transcript  that leads to a {\em maximal} production rate.
It is clear that a very small density will not be optimal, and since
the ribosomes interact and may jam each other, a very high density is also not optimal.

We studied this question using dynamical models for ribosome flow
in both a linear and a  circular mRNA molecule. Our results show that typically the
optimal density is close to one half of the maximal density.

In synthetic biology and biotechnology optimizing the translation rate is a standard goal, and we believe that our results
can provide guidelines for designing and reengineering transcripts.
However, in vivo biological regulation  of mRNA translation may have several goals besides optimizing the production rate.
For  endogenous genes there are many additional constraints that shape the transcript, translation rates, and ribosome densities. For example, it is known that evolution optimizes not only protein levels, but also attempts to minimize
their production cost~\cite{Tuller2010,Li2014}. This cost may include for example the
biocellular budget required for producing the ribosomes  themselves. Thus,
we do not expect that the protein levels of all genes will be maximal.
Rather, we expect that  translation is optimized for
proteins that are required with
 high copy numbers (e.g. those related to
house keeping genes and some structural genes).

Furthermore, it is important to mention that there are various additional constraints shaping the coding regions of endogenous genes. These   include  various regulatory signals related to various gene expression steps, co-translational folding, and the functionality of the protein \cite{Tuller2015,Zafrir2015,Zhang2009,Pechmann2013,Cartegni2002,Stergachis2013}. Thus, under these additional constraints we do not necessarily expect to see
 ribosome densities that maximize the  translation rate.

Indeed, experimental studies of ribosome densities in various
organisms
  demonstrate that on average $15\%-20\%$ of the
mRNA is occupied by ribosomes~\cite{Arava2003,Piques2009}.
 However, in~$241$ genes in 
 {\em S. cerevisiae} 
  more than $40\%$ of the mRNA is occupied by ribosomes~\cite{Arava2003}.
 This suggests that  a   ribosome density that is close to $0.5$ is   frequent in certain
\emph{specific}  mRNA molecules. In addition, it seems
 that under stress conditions ribosomal densities (and traffic jams) increase
 (see, e.g. \cite{Subramaniam2014}). Thus,
under such conditions we expect  more mRNAs with  ribosome densities close to $0.5$ (see, for example, \cite{Picard2013}).

Interestingly, the reported results are also
in   agreement with genome-wide simulations of the RFM that were performed based on the modeling of all
 the endogenous genes of {\em S. cerevisiae}, as reported in~\cite{reuveni}. Indeed,   Fig.~4C there shows
the ribosome density,  averaged over all the sites of all the mRNAs, as a function of the initiation rate.
 The maximal production rate corresponds
to an average density of about~$0.5$.

We note in passing that for an RFM with dimension $n$, with all the rates  equal (i.e. $\lambda_0=\cdots=\lambda_n$), the average ribosomal density 
is~$1/2$ for all~$n$, and that for an RFM with dimension~$n$, $\lambda_0\to\infty$, and equal elongation rates (i.e. $\lambda_1=\cdots=\lambda_n$), the average ribosomal density is $\frac{n+1}{2n}$, thus approaching $1/2$ as $n$ increases~\cite{rfm_wireless}.

Further studies
 may consider optimizing the translation rate
 under  various additional constraints. For example, it will be interesting to study the optimal ribosome density
when taking into account also
 the biocellular cost of protein production, or under given
 constraints on the allowed density profile, etc. In addition, it will be interesting to study the optimal densities in more comprehensive models that include competition for the free ribosomes between several
mRNA molecules~\cite{RFM_model_compete_J}.
Another important issue, that is not captured by the RFM and RFMR, is that every ribosome covers several codons.
Developing and analyzing  RFM/RFMR models with ``extended objects''
 is an important challenge.

Finally, TASEP has been used to model and analyze many other natural and artificial processes including traffic flow and the movement of motor proteins. The problem of the optimal density is of importance in these applications as well.

%%%%%%%%%%%%%%%%%%%%%%%%%%%%%%%%%%%
\section*{Acknowledgments}
%%%%%%%%%%%%%%%%%%%%%%%%%%%%%%%%%%%
We thank Gilad Poker for helpful comments.

%%%%%%%%%%%%%%%%%%%%%%%%%%%%%%%%%%%
\section*{Appendix: Proofs}
%%%%%%%%%%%%%%%%%%%%%%%%%%%%%%%%%%%
{\sl Proof of Proposition~\ref{prop:rfmr_dense}.}
%%%%%%%%%%%%%%%%%%%%%%%%%%%%%%%%%%%%%%%%%%%%%%%%%%%%%%5
It follows from known results on the solutions of ODEs that~$e_i$ is continuous in~$s$ for all~$i$.
It is known that every~$e_i$ is strictly increasing  in~$s$~\cite[Theorem 1]{rfmr}.
 Hence, there exists a set~$E$ of measure zero such that for all~$i$ and all~$s\in[0,n]\setminus E$
the derivative
$e_i':=\frac{d}{ds}e_i$ exists, and is strictly positive.
%%%%%%
The steady-state production rate satisfies~$R= \lambda_i e_i( 1 - e_{i+1})$, for all~$ i=1,\dots,n$. This yields
\be\label{eq:rfmr_r_deriv}
R'=\lambda_i( e_i'(1-e_{i+1})-e_ie_{i+1}'),
\ee
for all~$i$ and all~$s\in[0,n]\setminus E$.

Let~$\sgn(\cdot):\R\to\{-1,0,1\}$ denote the sign function, i.e.
\[
\sgn(y)=\begin{cases}1,& y>0,\\0,&y=0,\\-1,&y<0.\end{cases}
\]
%%%
We require the following result.
\begin{Proposition}\label{prop:rfmr_R'sign}
%%%%%%%%%%%%%%%%%%%%%%%%%%
For any~$s\in[0,n]\setminus E$,
\[
\sgn(R')=\sgn(\prod_{i=1}^n (1-e_i)-\prod_{i=1}^n e_i ).
\]
\end{Proposition}

{\sl Proof of Proposition~\ref{prop:rfmr_R'sign}.}
%%%%%%%%%%%%%%%%%%%%%%%%%%%%
Assume that~$R'>0$. Then~\eqref{eq:rfmr_r_deriv} yields
\[
e_i'(1-e_{i+1})>e_ie_{i+1}', \quad i=1,\dots,n.
\]
Multiplying   these~$n$ inequalities, and using the fact that~$e'_i>0$ for all~$i$  yields
\be\label{eq:rfmr_ei1>ei}
\prod_{i=1}^n (1-e_i) > \prod_{i=1}^n e_i.
\ee

To prove the converse implication, assume that~\eqref{eq:rfmr_ei1>ei} holds.
 Multiplying both sides of the inequality by the strictly positive term $\prod_{j=1}^n e_i'$ yields
\[
\prod_{i=1}^n a_i > \prod_{i=1}^n b_i,
\]
where $a_i:=e_i'(1-e_{i+1})$, and $b_i:=e_i e_{i+1}'$. This means that $a_\ell > b_\ell$ for some index $\ell\in\{1,\dots,n\}$.
Since $R'=\lambda_\ell(a_\ell-b_\ell)$, it follows that $R'>0$. Thus, we showed that $R'>0$ if and only if $\prod_{i=1}^n (1-e_i) > \prod_{i=1}^n e_i$. The proof that $R'<0$ if and only if $\prod_{i=1}^n (1-e_i) < \prod_{i=1}^n e_i$ is similar. This implies that $R'=0$ if and only if $\prod_{i=1}^n (1-e_i) = \prod_{i=1}^n e_i$, and this completes the proof of Proposition~\ref{prop:rfmr_R'sign}.~\IEEEQED

We can now complete the proof of  Proposition~\ref{prop:rfmr_dense}.
Let $p(s):=\prod_{i=1}^n (1-e_i)$, and $q(s):=\prod_{i=1}^n e_i$. Then  $p(0)=1$, $p(n)=0$, $q(0)=0$, and $q(n)=1$.
The strict monotonicity of every~$e_i$ implies  that $p(s)$ [$q(s)$] is a strictly decreasing [increasing] function in the interval $s\in[0,n]$.
This implies that there is a unique~$s^*\in[0,n]$  such that $p(s^*)=q(s^*)$.
By Proposition~\ref{prop:rfmr_R'sign}, this is the unique maximizer of~$R(s)$, and for~$s=s^*$:
\be\label{eq:prodsi}
				e_1^*\dots e_n^*=(1-e_1^*)\dots (1-e_n^*).
\ee
Also,
\begin{align}\label{eq:rstarloop}
%%%%%%%%%%%%%%%%%%%%%%%%%%%%%%%%%%%
									R^*&=\lambda_1 e_1^* (1-e_2^*)  \nonumber \\
									&=\lambda_2 e_2^* (1-e_3^*) \nonumber \\
									&\vdots\\
									&=\lambda_n e_n^* (1-e_1^*) \nonumber ,
%%%%%%%%									
\end{align}
and this yields~$(R^*)^n=(\lambda_1\dots\lambda_n)( e_1^*\dots e_n^*) ((1-e_1^*) \dots (1-e_n^*) )$.
Using~\eqref{eq:prodsi} completes the proof
 of Proposition~\ref{prop:rfmr_dense}.~\IEEEQED

{\sl Proof   of Fact~\ref{fact:sim}.}
%%%%%%%%%%%%%%%%%%%%%%%%%%%%%%%%%%%%%%%%%%%%%%%%%%%%%%%%%%%%%%%%%%
For~$n=2$, ~\eqref{eq:prodsi}   yields~$e^*_1+e^*_2=1$,
 and substituting this in~\eqref{eq:rstarloop} yields~\eqref{eq:rstar2}.
Consider the case~$n=3$. Let~$\lambda:=\lambda_1 \lambda_2\lambda_3$.
It follows from~\eqref{eq:rstarloop} that
\begin{align*}
%%%%%%%%%%%%%%%%%%%%%%%%%%%%%%%%%%%
							\lambda_2\lambda_3 		R^*&=\lambda e_1^* (1-e_2^* )   , \\
							\lambda_1\lambda_3 		R^*&=\lambda e_2^* (1-e_3^*)   , \\
							\lambda_1\lambda_2 		R^*&=\lambda e_3^* (1-e_1^*)   .
%%%%%%%%									
\end{align*}
Summing   these equations yields
\be\label{eq:n3so}
					\eta R^*=\lambda s^* -\lambda (  e_1^* e_2^* +  e_2^* e_3^* + e_3^* e_1^*   ),
\ee
where~$\eta:=\lambda_2\lambda_3+ \lambda_1\lambda_3 +	\lambda_1\lambda_2  $. It follows from~\eqref{eq:prodsi}  that
\[
						e_1^* e_2^* +  e_2^* e_3^* + e_3^* e_1^* = s^*-1+2 e_1^* e_2^* e_3^*,
\]
and substituting this in~\eqref{eq:n3so} yields~$\eta R^* = \lambda(1-2  e_1^* e_2^* e_3^*)$. Applying~\eqref{eq:rnstar}
completes the proof.~\IEEEQED

{\sl Proof of Proposition~\ref{prop:rfmr_sense}.}
%%%%%%%%%%%%%%%%%%%%%%%%%%%%%%%%%%%%%%%%%%%%%%%%%%%%
Write~\eqref{eq:rfmr_r_deriv} as
\be\label{eq:cd}
				D (e^*)'=C(e^*)',
\ee
where~$D:=\diag( 1-e^*_2,1-e^*_3,\dots,1-e^*_n,1-e^*_1 )$, and
\[
C:=\begin{bmatrix}
 0& e_1^* &0 &0 &\dots & 0 &0 \\
 0& 0&    e_2^* &0  &\dots & 0 &0 \\
&&\vdots\\
0& 0&    0 &0  &\dots & 0 &e^*_{n-1} \\
e^*_n& 0&    0 &0  &\dots & 0 &0
\end{bmatrix}.
\]
%%%%%
Note that~$C$ is cyclic of order~$n$, so multiplying~\eqref{eq:cd} by~$C^{n-1}$ yields
\be\label{eq:cdh}
				H (e^*)'=  (e_1^* \dots e_n^*) (e^*)',
\ee
where~$H:=C^{n-1}D$.
In other words,~$(e^*)'$ is an eigenvector of~$H$ corresponding to the eigenvalue~$(e_1^* \dots e_n^*)$.
The cyclic structure of~$C$ implies that
\begin{align*}
 C^{n-1}
=\begin{bmatrix}
 0& 0     &\dots & 0 &\mu^*_1 \\
 \mu^*_2& 0    &\dots & 0 &0 \\
&&\vdots\\
0   &\dots & \mu^*_{n-1}& 0 &0 \\
0& 0&     \dots & \mu^*_n &0
\end{bmatrix} ,
\end{align*}
where~$\mu_i^*:=e^*_i e^*_{i+1}\dots e^*_{i+n-2}$, with all indexes
interpreted modulo~$n$ (e.g.,~$e^*_{n+1}=e^*_1$). Now it is straightforward to verify that~$(e^*)'=c v$, with~$c\not =0$,
 is the only solution   of~\eqref{eq:cdh}.
Since every~$e_i$ increases with~$s$, we conclude that~$c>0$.
Furthermore,~$\sum_{i=1}^n e^*_i=s$ implies that~$\sum_{i=1}^n (e^*_i)'=1$, and this completes the proof.~\IEEEQED

%%%%%%%%%%%%%%%%%%%%%

{\sl Proof   of Proposition~\ref{prop:rfmr_half}.}
The proof
 follows immediately from the following
 result.
%%%%%%%%%%%%%%%%%%%%%%%%%%%%%%%%%%%%%%
\begin{Proposition}\label{prop:rfmr_maxR}
Consider an RFMR with dimension $n$,
 and  suppose that the transition rates satisfy~$\lambda_i=\lambda_{n-i}$ for all~$i$.  Then
\begin{enumerate}
\item $e^*_i=e^*_{n+1-i}$ for any~$i$;
\item $R(s)=R (n-s)$ for any~$s\in[0,n]$, and
   $R(s_1) < R(s_2)$ for any~$0\le s_1<s_2\leq n/2$.
\end{enumerate}
\end{Proposition}
This  means in particular that $R(s)$ is symmetric with respect to $s=n/2$, and is strictly increasing in the interval $[0,n/2)$.

{\sl Proof of Proposition~\ref{prop:rfmr_maxR}.}
%%%%%%%%%%%%%%%%%%%%%%%%%%%%%%%%%%%%%%%%%%%%%%%%%%%%%%%
Given an RFMR with dimension $n$, and rates~$\lambda_i$, $i=1,\dots,n$,
let $\bar x_i(t):=1-x_{n+1-i}(t)$, $i=1,\dots,n$.
Then using the equation
\[
\dot{x}_i=\lambda_{i-1}x_{i-1}(1-x_i)-\lambda_i x_i(1-x_{i+1})
\]
yields
\[
\dot{\bar x}_i=\bar \lambda_{i-1}\bar x_{i-1}(1-\bar x_i)-\bar \lambda_{i} \bar x_i(1-\bar x_{i+1}) ,
\]
with~$\bar \lambda_i:=\lambda_{n-i}$ (recall that all indexes are interpreted modulo~$n$).
This is again an RFMR. Fix an arbitrary~$s\in[0,n]$. Then for any~$x(0)$ such that~$1_n^T x(0)=s$
we have~$1_n^T \bar x(0)=n-s$. Therefore,
the~$x$ system converges to~$e=e(s,\lambda_1,\dots,\lambda_n)$, and the~$\bar x$ system to~$\bar e=e(n-s,\bar
\lambda_1,\dots,\bar \lambda_n)$. This implies that~$e_i(s,\lambda_1,\dots,\lambda_n)=1-e_{n+1-i}(n-s,\bar \lambda_{1},\dots,\bar \lambda_n)$ for all~$i$. The steady-state production rate in the~$\bar x$ system is
%%%%%%%%%%%%%%%%%%%%%%%
\begin{align*}
  \bar R&= \bar \lambda_n   \bar e_n (1 - \bar e_1) \\
        &=\lambda_n (1-e_1 ) e_n \\&=R.
\end{align*}
If the rates satisfy~$\lambda_i=\lambda_{n-i}$ for all~$i$ then~$e_i(s)=1-e_{n+1-i}(s)$ for all~$i$,
 and~$R(s)=R(n-s)$. By Proposition~\ref{prop:rfmr_dense}, this means that~$R^*=R(n/2)$.
Combining this with the results in the
proof of Proposition~\ref{prop:rfmr_dense} completes the proof of Proposition~\ref{prop:rfmr_maxR}.~\IEEEQED

%%%%%%%%%%%%%%%%%%%%%%%%%%%%

{\sl Proof of Proposition~\ref{prop:lambda_ratio_hom}.}
%%%%%%%%%%%%%%%%%%%%%%%%%%%%%%%%%%%%%%%%%%%%%%%%%%%%%%
Consider Problem~\ref{prob:max} and the homogeneous constraint~\eqref{eq:homog_comnst}.
By~\cite[Proposition $4$]{rfm_dyson}:
\be\label{eq:e_sym}
e_i^*=1-e_{n-i+1}^*,
\ee
and
\be\label{eq:g_e_opt}
\frac{\lambda_{i}^*}{\lambda_{i-1}^*}= \frac{e_{i}^*}{1-e_{i}^*},
\ee
$i=1,\dots,n$, and by~\cite[Theorem $1$]{rfm_dyson}:
\be\label{eq:l_dec}
\lambda_0^*<\lambda_1^*<\dots<\lambda_{\lfloor n/2 \rfloor}^*,
\ee
and
\be\label{eq:l_sym}
\lambda_i^*=\lambda_{n-i}^* ,\quad  i=0,\dots,n.
\ee
Thus,~\eqref{eq:e_sym} proves~\eqref{eq:eismono}, and combining~\eqref{eq:l_dec},~\eqref{eq:l_sym}, and~\eqref{eq:g_e_opt} yield~\eqref{eq:eisneven} and~\eqref{eq:eisnodd}.~\IEEEQED

%%%%%%%%%%%%%%%%%%%%%%%%%%%%%%%%%%%%%%%%%%%%%%%%%%%%%%%%%%%%
\bibliographystyle{IEEEtranS}
\bibliography{RFM_bibl_for_opt_dense,rfm_kdown}

\end{document}